\documentclass[epj]{svjour}
\usepackage{graphics}
\def\Re{{\rm Re\mit}}
\def\Im{{\rm Im\mit}}
\begin{document}
\title{Influence of the sign of the refractive index in the reflectivity of a metamaterial surface with localized roughness}
\author{Vivian Gr\"{u}nhut \thanks{vgrunhut@df.uba.ar}
\and Ricardo A. Depine \thanks{rdep@df.uba.ar}
}                     
%
%
\institute{Grupo de Electromagnetismo Aplicado, Departamento de F\'{\i}sica, 
Facultad de Ciencias Exactas y Naturales, Universidad de Buenos Aires, Ciudad Universitaria Pabell\'{o}n I, C1428EHA, 
Buenos Aires, Argentina 
}
\date{Received: date / Revised version: \today}
%
\abstract{To study the scattering properties of metamaterials, we generalize two scattering methods developed for conventional (non-magnetic) isotropic materials to the case of materials with arbitrary values (positive or negative) of magnetic permeability and electric permittivity. The generalized methods are used to study the changes 
in the reflectivity of a metamaterial surface with localized roughness when the relative refractive index changes sign. Our results show that, unlike the case of a plane surface whose reflectivity is unaffected by the change of sign of the relative refractive index, in rough surfaces the change of sign is manifested in the reflectivity, even for very low roughness, particularly in observation directions away from the specular direction. 
} 
\authorrunning{Gr\"unhut and Depine}
\titlerunning{Influence of the sign of the refractive index in the reflectivity of a metamaterial with roughness}
\maketitle
\section{Introduction}
\label{intro}

The area of metamaterials has grown rapidly during the first decade of the twenty-first century. While there is still no consensus on the definition of the term \emph{metamaterial}, a short and broad definition could be: 
an artificial environment with electromagnetic properties nonexistent or very difficult to find in natural materials \cite{MMs1,MMs2,MMs3}. Among these properties, the one which has perhaps attracted more attention from the scientific community is the negative refractive index \cite{NIM2}. In the ideal case of lossless media, negative refractive index occurs when there is a range of frequencies over which both the electric permittivity $\epsilon$ and the magnetic permeability $\mu$ are simultaneously negative \cite{NIMVese}. For this reason many authors refer to materials with negative index as materials with negative constitutive parameters \cite{MMs2}. For the actual case of media with losses, the condition of negative refractive index is wider \cite{npv2} and can be written as
\begin{equation}
\epsilon_{R}\,|\mu|+\mu_{R}\,|\epsilon|<0, \label{npvppv}
\end{equation}
where $\epsilon_{R}=\Re\,\epsilon$, $\mu_{R}=\Re\,\mu$ and $\Re$ indicates the real part of a complex quantity. 
It has been recently suggested that
this condition is only valid for passive media \cite{Lak0}.

While in a material with positive refractive index the vectors electric field $\vec{E}$, magnetic field  $\vec{H}$ and direction of propagation $\vec{k}$ of a plane wave form a right 
handed set, in a material with negative refractive index these vectors form a left handed set. 
This is why some authors use the terms right-handed (RH) and left-handed (LH) to refer to materials with positive and negative refractive index respectively, even though we think it would be more appropriate to reserve such designations for chiral media \cite{quirales}, where the internal structure of the medium is associated with a direction of rotation. As a result of the positive or negative character of the triplet $\vec{E}$, $\vec{H}$, $\vec{k}$, the Poynting vector $\vec{S}$ becomes parallel or antiparallel to the wave vector $\vec{k}$. Therefore, conventional materials are also referred to as positive phase velocity (PPV) media whereas materials with negative refractive index are referred to as negative phase velocity (NPV)  media \cite{npv2,npv1}. 

The seemingly simple change of sign of the refractive index produces dramatic changes in well-known phenomena as the Doppler effect, the law of refraction or Cerenkov radiation \cite{NIMVese}. In this paper we are interested in investigating the changes in the scattering properties of a non-periodic rough surface that separates a conventional medium from a metamaterial when the refractive index of the metamaterial changes sign. Similar problems have already been discussed in the following cases: i) limited volumes of simple shape, such as cylinders \cite{cil1,cil2}, spheres \cite{esf1,esf2} and ii) gratings formed by isotropic \cite{rdep1,rdep2,rdep3,rdep4} and uniaxial media \cite{rdep5,rdep6}. However, to our best knowledge there are no similar studies so far for the paradigmatic case of two isotropic half-spaces separated by a non-periodic rough surface. 
Studies of this type may be relevant not only in novel applications where the properties of metamaterials play a crucial role, for example in the design of invisibility cloaks \cite{invis1,invis2}, perfect lenses \cite{lens1}, control of the Casimir force \cite{casimir1} or excitation of surface polaritons \cite{mauro1,mauro2},  but also in more conventional applications, similar to those used for non-magnetic media, such as determining the constitutive parameters of a metamaterial from experimental reflectance curves as a function of the incidence angle \cite{retrieval1}.
Besides the interest motivated by the mentioned applications, investigating the influence of roughness on the electromagnetic response of almost flat surfaces that differ only in the sign of the refractive index on both sides of the surface may also reveal features that could be used in nondestructive analysis techniques to distinguish the PPV and NPV character of a given rough surface. This is so because for perfectly flat surfaces between lossless media, reflectance curves as a function of incidence angle $\theta_{0}$ do not distinguish between PPV and NPV media with the same absolute value of the relative refractive index. This property is part of a more general conjugation symmetry \cite{Lakconjuga}, valid even for lossy refracting media, which ensures that the transformation 
\begin{equation}
\big\{\epsilon\rightarrow -\epsilon^*,\mu\rightarrow -\mu^*\big\},  \label{conjugation}
\end{equation}
(where $\epsilon$ and $\mu$ now represent the relative parameters and the asterisk denotes the complex conjugate) changes the phase but not the magnitude of the Fresnel coefficients for the amplitudes of reflected and transmitted fields. 
This conjugation symmetry is only valid for non-evanescent incident waves, i.e, for real angles of incidence $\theta_{0}$ and with $0 \le |\theta_{0}| \le \pi/2$. 
Taking into account that the presence of roughness introduces non specular components in the total fields generated by an illuminated boundary, 
the symmetry mentioned before suggests that a far field indicator of the PPV or NPV character of a rough surface could be found in observation directions away from the specular direction. 
Alternatively, because the presence of roughness introduces evanescent (non radiative) components 
in the total fields generated by an illuminated boundary, this conjugation symmetry, together with the fact that evanescent waves behave in opposite ways in PPV and in NPV media \cite{NIMVese,lens1}, indicates that the PPV or NPV character of a rough surface should also be revealed in near field observations.

In order to explore these issues in an electromagnetically rigorous way, numerical treatments are inevitable and it is convenient and desirable to develop efficient and simple methods 
to evidence the physical mechanisms involved in the interaction between the incident wave, the rough surface and the properties of the refracting material, 
while preventing the physical mechanisms from being masked by the numerical treatments. 
Since it is not generally easy to combine both features in multi-purpose methods, such as finite element or finite-difference time-domain methods where Maxwell's equations are discretized from the beginning, in this work we present two relatively simple treatments that meet the above conditions and which are based on what is known as the Rayleigh hypothesis \cite{Rayleigh1}. This hypothesis, used by Lord Rayleigh in 1907 to solve the dispersion of an acoustic wave by an impenetrable periodic surface, states that the fields near the corrugation are composed of waves moving away from the surface, an assumption that was objected by Lippmann \cite{Lippmann} in 1953. Lippmann's objection led to numerous studies devoted to establishing the limit of the validity of the Rayleigh hypothesis (for an historical overview see for example \cite{RayN0})).  
The application of this hypothesis in various configurations continues to arouse interest, as shown in references \cite{RayN1,RayN2,RayN3,RayN4,RayN5}. 
Nowadays, it is recognized that in the case of conventional materials the Rayleigh hypothesis gives good results for low roughness surfaces. The same has been verified in the case of metamaterials with negative refractive index, where the Rayleigh hypothesis has been successfully employed to study diffraction from periodically corrugated surfaces \cite{rdep1,rdep2,rdep3,rdep5}. In this paper we use the Rayleigh hypothesis to investigate the changes in the scattering properties of a non-periodic rough surface that separates a conventional medium from a metamaterial, when the refractive index of the metamaterial changes sign. 
In Section \ref{sec:1} we provide a brief 
description of the boundary value problem 
for the scattering of a plane wave at a rough metamaterial surface, obtaining a system of coupled integral equations for the amplitudes of the scattered fields on both sides of the surface. 
Next, we decouple the system of integral equations in order to obtain the generalization for magnetic media of the reduced Rayleigh equations (previously obtained by Toigo et al. \cite{Toigo} for conventional materials) and outline two methods of resolution that allow us to calculate the reflected and transmitted fields in an independent way: a direct numerical method, limited only by the validity of the Rayleigh hypothesis, and a perturbative method, valid when the height of the corrugation is small compared to the wavelength of the incident radiation. The perturbative method, originally developed by Rice \cite{Rice} for impenetrable media, 
leads to a relatively simple numerical treatment and is very useful to validate the results obtained with the direct numerical method. 
Section \ref{numerical} is devoted to discussing the numerical results obtained with both methods for the case of deterministic surfaces with a single corrugation, 
postponing the study of statistically characterized rough surfaces for future work. 
We use a time dependence of the type $e^{-i\omega t}$ where $\omega$ is the angular frequency, $t$ the time and $i=\sqrt{-1}$.

\section{Analysis}
\label{sec:1}
\subsection{The boundary value problem}
\label{sec:2}
%
\begin{figure}
\resizebox{0.5\textwidth}{!}{%
\includegraphics{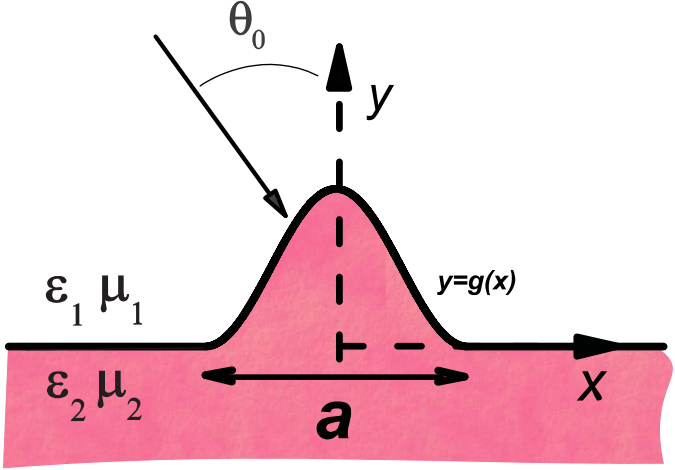}
}
\caption{Outline of the boundary value problem.}
\label{fig1}       
\end{figure}
Consider a rough surface represented by the function $y=g(x)$ (see Fig. \ref{fig1}). This surface separates two homogeneous and  isotropic materials characterized by the constitutive parameters $\epsilon_{i}$ (electric permittivity) and $\mu_{i}$ (magnetic permeability), $i=1,2$. Medium 1 ($y>g(x)$, medium of incidence) is a conventional material with positive refractive index $\nu_1=\sqrt{\epsilon_{1}\mu_{1}}$, $\epsilon_{1}>0$,  $\mu_{1}>0$, while medium 2 ($y<g(x)$, refractive media) is a metamaterial with frequency-dependent constitutive parameters $\epsilon_{2}=\epsilon_{2R}+i\epsilon_{2I}$ and $\mu_{2}=\mu_{2R}+i\mu_{2I}$, 
real parts $\epsilon_{2R}$ and $\mu_{2R}$ of arbitrary sign and positive imaginary parts $\epsilon_{2I}>0$ and $\mu_{2I}>0$.  
The metamaterial, with refractive index $\nu_2=\sqrt{\epsilon_{2}\mu_{2}}$, is either an NPV material, if its  constitutive parameters satisfy eq.~(\ref{npvppv}), or a PPV material otherwise. The rough surface is illuminated by an electromagnetic, linearly polarized plane wave that propagates in the $(x,y)$ plane (incidence plane) and forms an angle $\theta_{0}$, $\left(|\theta_{0}|<\pi/2\right)$ with the $y$ axis. 
We analyze two independent polarization cases separately: the $s$ or TE polarization (electric field in the $z$ direction) and the $p$ or TM polarization (magnetic field in the $z$ direction). In both cases, the scattered fields 
(reflected and transmitted) conserve the polarization of the incident wave. 
We denote by $\Psi(x,y)$ the $z$-directed component either of the total electric field ($s$ polarization) or 
the total magnetic field ($p$ polarization). It is known~\cite{campo} that outside the corrugated region ($\min g(x)\leq y\leq\max g(x)$), $\Psi(x,y)$ can be rigorously represented by superpositions of plane waves. If $y > \max g(x)$,
\begin{eqnarray}
\Psi_1(x,y)&=&e^{i\left(\alpha_{0}x-\beta_{0}^{(1)}y\right)} \nonumber \\ 
&+& \frac{1}{2\pi}\int_{-\infty}^{+\infty} R(\alpha){e^{i\left(\alpha x+\beta^{(1)}_{\alpha}y\right)}}d\alpha 
\label{ecu3}
\end{eqnarray}
represents the incident plane wave (first term, with unit amplitude) and the scattered fields in medium 1 (second term, reflected field), while if $y <\min g(x)$,
\begin{equation}
\Psi_2(x,y)=\frac{1}{2\pi}\int_{-\infty}^{+\infty} 
T(\alpha){e^{i\left(\alpha x-\beta^{(2)}_{\alpha}y\right)}}d\alpha 
\label{ecu4}
\end{equation}
represents the scattered fields in medium 2 (transmitted field). 
The quantity $\alpha_{0}=k_0 \nu_1 \sin \theta_{0}$, $k_0=\omega/c$, represents the $x$ component of the incident wave vector. 
Note that the integrand in (\ref{ecu3}) represents a plane wave with amplitude $R(\alpha)$ and wave vector 
$\vec k^{(1r)}(\alpha)=\alpha \hat x +\beta^{(1)}_{\alpha} \hat y$, while the integrand in (\ref{ecu4}) represents a plane wave with amplitude $T(\alpha)$ and wave vector $\vec k^{(2t)}(\alpha)=\alpha \hat x -\beta^{(2)}_{\alpha} \hat y$. 
The components along $y$ of the wave vectors $\vec k^{(1r)}$ and $\vec k^{(2t)}$ are
\begin{equation}
\beta^{(j)}_{\alpha}=\beta^{(j)}(\alpha)=\left(k_0^2 \epsilon_{j}\mu_{j}-\alpha^2\right)^{1/2}, \,\,\,\,j=1,2, 
\label{ecubetas}
\end{equation}
and we define $\beta_{0}^{(j)}=\beta^{(j)}(\alpha_{0})$. Also, note that the quantities $\beta^{(1)}_{\alpha}$ are real
or purely imaginary. In the first case, which occurs in the so-called radiative zone $|\alpha/k_0| < \nu_{1}$, we must require 
fulfillment of the condition $\Re \, \beta^{(1)}_{\alpha}\geq 0$, in order that the fields in eq. (\ref{ecu3}) represent propagating plane waves that move away from the surface into the half-space $y>g(x)$. In the second case, which occurs in the so-called non radiative zone $|\alpha/k_0| \ge \nu_{1}$, we must require fulfillment of the condition 
$\Im \, \beta_{\alpha}^{(1)} \geq 0$, in order that these fields represent evanescent waves that attenuate for $y\rightarrow +\infty$. Similar considerations are valid for the quantities 
$\beta^{(2)}_{\alpha}$ in the ideal case of completely transparent transmission media (lossless, $\epsilon_{2}$ y $\mu_{2}$ real), although it should be noted that in this case the choice of the branches of the square root function in (\ref{ecubetas}) depends on the PPV or NPV character of medium 2. 
%
On the other hand, in the real case of lossy transmission media ($\Im \, \epsilon_{2}>0$, $\Im \,\mu_{2} >0$), 
the quantities $\beta_{\alpha}^{(2)}$ are always complex with a nonzero imaginary part, 
$\Im \,\beta^{(2)}_{\alpha}>0$, in order that the fields in eq. (\ref{ecu4}) attenuate for $y\rightarrow-\infty$. 
Note that the condition $\Im \, \beta_{\alpha}^{(2)}>0$ automatically sets the sign of $\Re \, \beta^{(2)}_{\alpha}$, independently of the signs of $\epsilon_{2R}$ and of $\mu_{2R}$, i.e., 
independently of the PPV or NPV character of medium 2. 

In order to obtain the unknown amplitudes $R(\alpha)$ and $T(\alpha)$ the appropriate boundary conditions at $y=g(x)$ must be imposed. These conditions can be written in the following way
\begin{eqnarray}
\Psi_1\left(x,g(x)\right)&=&\Psi_2 \left(x,g(x)\right), \label{cc1}\\
\frac{1}{\sigma_{1}}\frac{\partial}{\partial \hat{n}}\Psi_1\left(x,g(x)\right)&=&
\frac{1}{\sigma_{2}}\frac{\partial}{\partial \hat{n}}\Psi_2\left(x,g(x)\right), \label{cc2}
\end{eqnarray}
where $\sigma_{j}=\mu_{j}$ for the $s$ mode or $\sigma_{j}=\epsilon_{j}$ for the $p$ mode and $\hat{n}$ is the unit vector normal to the surface. 

\subsection{Rayleigh-hypothesis}
\label{sec:2}
In order to satisfy the boundary conditions (\ref{cc1}) and (\ref{cc2}) we use the Rayleigh~\cite{Rayleigh1} hypothesis, i.e.,  we assume that the equations (\ref{ecu3}) and (\ref{ecu4}), that strictly represent the fields outside the corrugated zone $\min g(x)\leq y\leq\max g(x)$, can also be used to represent the fields near the surface. Proceeding in this way and projecting the boundary conditions in the base of Rayleigh functions $\{e^{i\alpha' x}\}_{\alpha'\epsilon\Re}$, we obtain a system of two coupled integral equations, whose unknowns are the complex amplitudes $R(\alpha)$ and $T(\alpha)$. 
%
This system can be decoupled through a procedure similar to those presented in Refs. \cite{Toigo,Lester}, thus obtaining one integral equation for the unknown amplitudes $R(\alpha)$ 
\begin{equation}
-K(\alpha_{0},\alpha)=\int_{-\infty}^{+\infty}K^{r}(\alpha^{\prime},\alpha)R(\alpha^{\prime})d\alpha^{\prime}\label{refl1}
\end{equation}
and another integral equation for the unknown amplitudes $T(\alpha)$ 
\begin{equation}
-2 \beta_{0}^{(1)}\ \frac{\sigma_{2}}{\sigma_{1}}\ \delta\left(\alpha-\alpha_{0}\right)=
\int_{-\infty}^{+\infty}K^{t}(\alpha^{\prime},\alpha)T(\alpha^{\prime})d\alpha^{\prime}\label{tran1}
\end{equation}
where $\delta\left( \right)$ is the Dirac delta distribution. The equations (\ref{refl1}) and (\ref{tran1}) are Fredholm integral equations of the first kind, with kernels
\begin{equation}
\begin{array}{ll}
K^{r}(\alpha^{\prime},\alpha)=M_{\alpha^{\prime},\alpha}D\left[\alpha-\alpha^{\prime},\beta^{(2)}_{\alpha}-\beta^{(1)}_{\alpha^{\prime}}\right]\label{ecu1}
\end{array}
\end{equation} and
\begin{equation}
\begin{array}{ll}
K^{t}(\alpha^{\prime},\alpha)=M_{\alpha,\alpha^{\prime}}D\left[\alpha-\alpha^{\prime},\beta^{(2)}_{\alpha^{\prime}}-\beta^{(1)}_{\alpha}\right]
\label{ecu2}
\end{array}
\end{equation}
where
\begin{equation}
\begin{array}{ll}
M_{\alpha^{\prime},\alpha}=\frac{\left(1-\frac{\sigma_{2}}{\sigma_{1}}\right)\left(\alpha\alpha'+\beta^{(2)}_{\alpha}\beta^{(1)}_{\alpha^{\prime}}\right)+k_0^2 \left(\frac{\sigma_{2}}{\sigma_{1}}\nu_{1}^2-\nu_{2}^2\right)}{\beta^{(2)}_{\alpha}-\beta^{(1)}_{\alpha^{\prime}}}
\end{array}
\end{equation}
and
\begin{equation}
D[u,v]=\frac{1}{2\pi  }\int_{-\infty}^{+\infty}dx e^{-iux}e^{-ivg(x)}\label{fourier}
\end{equation}
is the Fourier transform of $e^{-ivg(x)}$.
The inhomogeneity in eq. (\ref{refl1}) is given by $K(\alpha_{0},\alpha)$, with  
\begin{equation}
\begin{array}{ll}
K(\alpha_{0},\alpha)=N_{\alpha_{0},\alpha}2\pi D\left[\alpha-\alpha_{0},\beta^{(2)}_{\alpha}+\beta_{0}^{(1)}\right]\label{inhom1}
\end{array}
\end{equation}
where
\begin{equation}
\begin{array}{ll}
N_{\alpha_{0},\alpha}=\frac{\left(1-\frac{\sigma_{2}}{\sigma_{1}}\right)\left(\alpha_{0}\alpha-\beta^{(2)}_{\alpha}\beta_{0}^{(1)}\right)+k_0^2 \left(\frac{\sigma_{2}}{\sigma_{1}}\nu_{1}^2-\nu_{2}^2\right)}{\beta^{(2)}_{\alpha}+\beta_{0}^{(1)}}. 
\end{array}
\end{equation}

\subsection{Direct numerical method}
\label{sec:2}
To solve integral equations as (\ref{refl1}) numerically, we first use a quadrature scheme 
that allows us to approximate the integral as a linear combination of the values $R(\alpha_j)$ of the unknown function $R$ evaluated at the points of a grid $\lbrace \alpha_j \rbrace_{j=1}^{N_\alpha}$ of the independent variable $\alpha$. 
The numerical parameter $N_\alpha$ controls the grid density and will be determined by convergence criteria. Second, the approximated version of the identity (\ref{refl1}) is evaluated in the discrete points $\lbrace \alpha_j \rbrace_{j=1}^{N_\alpha}$,  thus obtaining $N_\alpha$ algebraic equations whose inversion will allow, in principle, to determine the unknowns $R(\alpha_j)$, $j=1, \ldots, N_\alpha$. The fact that the integration interval of the $\alpha$ variable in (\ref{refl1}) is infinite can be overcome by assuming that $|R(\alpha)| \rightarrow 0$ when $|\alpha|\rightarrow \pm \infty$. In this case, the integral over the infinite interval can be approximated by an integral over a finite interval $|\alpha| \le \alpha^{\mbox{Max}}$, where $\alpha^{\mbox{Max}}$ is another numerical parameter to be determined a posteriori through convergence criteria. 
%
A similar treatment for the equation (\ref{tran1}) allows us to determine $T(\alpha_j)$, the values of 
the unknown function $T$ evaluated at the points of the grid $\lbrace \alpha_j \rbrace_{j=1}^{N_\alpha}$. 

Taking into account that for a flat surface ($g(x)\equiv0$) the functions $R(\alpha)$ and $T(\alpha)$ are  proportional to Dirac delta distributions, we expect the functions $R(\alpha)$ and $T(\alpha)$ in the case of slightly corrugated surfaces to be highly concentrated around specular observation directions, that is $\alpha\approx \alpha_{0}$. Neither this feature nor the existence of a Dirac delta in eq. (\ref{tran1}) are convenient from a numerical point of view, but this can be overcome by introducing new functions $\tilde R(\alpha)$ and $\tilde T(\alpha)$ defined as
\begin{equation}
R(\alpha)=R^{(0)} \delta\left(\alpha-\alpha_{0}\right) + \tilde R(\alpha) \,,\label{redefinoR}
\end{equation}
\begin{equation}
T(\alpha)=T^{(0)} \delta\left(\alpha-\alpha_{0}\right) + \tilde T(\alpha) \,,\label{redefinoT}
\end{equation}
with $R^{(0)}$ and $T^{(0)}$ 
\begin{equation}
R^{(0)}=\frac{\frac{\sigma_{2}}{\sigma_{1}}\beta^{(1)}_{0}-\beta^{(2)}_{0}}{\frac{\sigma_{2}}{\sigma_{1}}\beta^{(1)}_{0}+\beta^{(2)}_{0}}\,,\label{fresnelref}
\end{equation}
\begin{equation}
T^{(0)}=2\frac{\frac{\sigma_{2}}{\sigma_{1}}\beta^{(1)}_{0}}{\frac{\sigma_{2}}{\sigma_{1}}\beta^{(1)}_{0}+\beta^{(2)}_{0}}\,, \label{fresneltra}
\end{equation}
the Fresnel coefficients for a perfectly flat surface. The new integral equations for the complex amplitudes $\tilde R(\alpha)$ and $\tilde T(\alpha)$ are
\begin{eqnarray}
-K(\alpha_{0},\alpha)= R^{(0)}(\alpha) \,K^{r}(\alpha_{0},\alpha)  \nonumber \\ 
+ \int_{-\infty}^{+\infty} K^{r}(\alpha^{\prime},\alpha)\tilde{R}(\alpha^{\prime})d\alpha^{\prime},
\label{refl1bis}
\end{eqnarray}
and
\begin{eqnarray}
-2\beta^{(1)}_{0}\frac{\sigma_{2}}{\sigma_{1}}= T^{(0)}(\alpha) K^{t}(\alpha,\alpha_{0}) \nonumber \\ 
+ \int_{-\infty}^{+\infty} K^{t}(\alpha,\alpha^{\prime})\tilde{T}(\alpha^{\prime})d\alpha^{\prime}. 
\label{tran1bis}
\end{eqnarray}

\subsection{Perturbative method}
\label{sec:2}
When the height of the corrugation is small compared to the incident wavelength $\lambda$, equations (\ref{refl1}) and (\ref{tran1}) can be solved by means of a standard \cite{Toigo,Rice} perturbative approach. To do so, we introduce the following power series expansions for $R(\alpha)$, $T(\alpha)$ and for terms of the form $e^{-ivg(x)}$ which appear in the kernels (\ref{ecu1}) and (\ref{ecu2}) and in the inhomogeneity (\ref{inhom1})
\begin{equation}
R(\alpha)=\sum_{n=0}^{\infty}\frac{R^{(n)}(\alpha)}{n!}\label{(des1)} \,,
\end{equation}
\begin{equation}
T(\alpha)=\sum_{n=0}^{\infty}\frac{T^{(n)}(\alpha)}{n!}\,,\label{(des2)}
\end{equation}
\begin{eqnarray}
e^{-ivg(x)}=\sum_{n=0}^{\infty}\frac{(-ivg(x))^n}{n!}\,. 
\end{eqnarray}
The integral (\ref{fourier}) can be written as
\begin{eqnarray}
D[u,v]=
\sum_{n=0}^{\infty}\frac{(-i)^nv^n}{n!}\hat{{g}}^{(n)}(u), \label{(des3)}
\end{eqnarray}
where $\hat{{g}}^{(n)}(u)$ is the Fourier transform of the function $[g(x)]^n$ and the index $n$ in the series (\ref{(des1)}), (\ref{(des2)}) and (\ref{(des3)}) indicates the perturbative order. 
When these expansions are introduced in the integral equations (\ref{refl1}) and (\ref{tran1}) 
the following iterative schemes for the coefficients $R^{(n)}(\alpha)$ and $T^{(n)}(\alpha)$, $n\geq 1$, are obtained 
\begin{eqnarray}
& R^{(n)}(\alpha)= \,\,\,\,\,\,\,\,\,\,\,\,\,\,\,\,\,\,\,\,\,\,\,\,\,\,\,\,\,\,\,\,\,\,\,\,\,\,\,\,\,\,\,\,\,\,\,\,\,\,\,\,\,\,\,\,\,\,\,\,\,\,\,\,\,\,\,\,\,\,\,\,\,\,\,\,\,\,\,\,\,\,\,\,\,\,\,\,\,\,\,\,\,\,\,\,\,\,\,\,\,\,\,\,\,\,\,\,\,\nonumber\\
& -  \bigg[(-i)^n2\pi\hat{{g}}^{(n)}(\alpha-\alpha_{0})N_{\alpha0}\left(\beta^{(2)}_{\alpha}+\beta^{(1)}_{0}\right)^n 
+  \nonumber\\
& \sum_{j=1}^{n}(-i)^j{n \choose j}\int_{-\infty}^{+\infty}d\alpha'M_{\alpha\alpha'}\left(\beta^{(2)}_{\alpha}-\beta^{(1)}_{\alpha'}\right)^j   \nonumber\\
& \hat{{g}}^{(j)}(\alpha-\alpha')R^{(n-j)}(\alpha')\bigg ]/\,M_{\alpha\alpha}\,, \label{pertur}
\end{eqnarray}
\begin{equation}
\begin{array}{ll}
T^{(n)}(\alpha)=-\bigg[\sum_{j=1}^{n}\\
(-i)^j{n \choose j}\int_{-\infty}^{+\infty}d\alpha'M_{\alpha'\alpha}\left(\beta^{(2)}_{\alpha'}-\beta^{(1)}_{\alpha}\right)^j\\\hat{{g}}^{(j)}(\alpha-\alpha')T^{(n-j)}(\alpha')\bigg]/M_{\alpha\alpha}\,,\label{pertut}
\end{array}
\end{equation}
where $R^{(0)}(\alpha)$ and $T^{(0)}(\alpha)$ coincide with the Fresnel coefficients given by the equations (\ref{fresnelref}) and (\ref{fresneltra}).

\section{Numerical treatment} \label{numerical}
The methods presented in the previous section have been implemented numerically for surfaces with a finite number of protuberances limited to region $-a/2\le x\le a/2$. To illustrate the changes produced in the reflectivity of a non flat surface when the sign of the relative index of refraction is changed, the incidence medium is vacuum ($\epsilon_{1}=1$, $\mu_{1}=1$) and the transmission medium is a lossy metamaterial in all examples presented here. The values of the constitutive parameters of the metamaterial are $\epsilon_{2}=5+0.01\,i$, $\mu_{2}=1+0.01\,i$ (a PPV medium with refractive index $\nu_{2}\approx 2.23+0.01\,i$) 
or $\epsilon_{2}=-5+0.01\,i$, $\mu_{2}=-1+0.01\,i$ (an NPV medium with refractive index $\nu_{2}\approx -2.23+0.01\,i$). 
Bear in mind that one set of constitutive parameters is obtained from the other set through the conjugation transformation expressed in eq. (\ref{conjugation}) and therefore both sets give the same reflectivity for a perfectly flat surface. 

To control the quality of the calculations, we have checked the convergence of the results for different values of the numerical parameters and the agreement between the direct and the perturbative methods. Besides, as energy is conserved in the scattering process, we have checked the fulfillment of the power conservation criterion. Taking into account that we are considering lossy media, it is convenient to write this criterion in the following form \cite{DeSanto}
\begin{equation}
P_r + P_a =1 \,,\label{conservation1}
\end{equation}
where 
\begin{equation}
P_r=\frac{\Re}{2\pi  } \int_{-\infty}^{+\infty}{\frac{\beta_{\alpha}^{(1)}}{\beta_{0}^{(1)}}|R(\alpha)|^2}\,d\alpha
\,,\label{conservation2}
\end{equation}
represents the fraction of the incident power which is scattered (reflected) into the incident medium, and 
\begin{equation}
\begin{array}{ll}
P_a={\frac{\sigma_{1}}{\sigma_{2}^{*}} \frac{\Re }{\beta_{0}^{(1)}}
\int_{-\infty}^{+\infty} 
\Bigl \lbrack
\Psi_{2}
\left(\frac{\partial \Psi_{2}^{*}}{\partial y} 
-\frac{\partial \Psi_{2}^{*}}{\partial x}g'\right)}\Bigr \rbrack_{y=g(x)}\,dx \,,\label{conservation3}
\end{array}
\end{equation}
represents the fraction of the incident power which is absorbed by the medium below the surface. 
%

\begin{figure}
\resizebox{0.5\textwidth}{!}{
\begin{tabular}{c}
\includegraphics{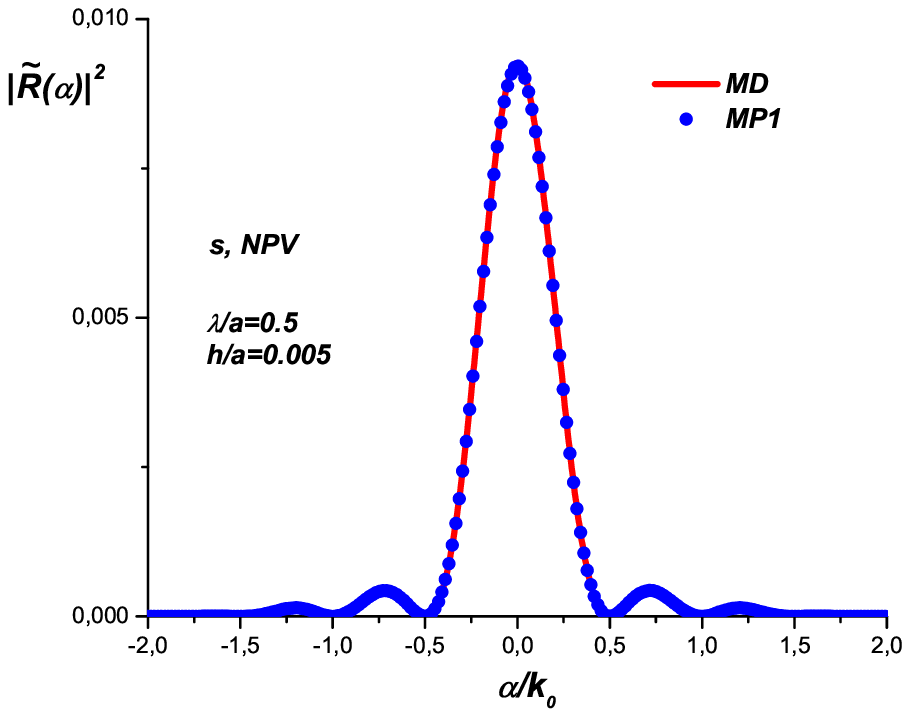}\\
\includegraphics{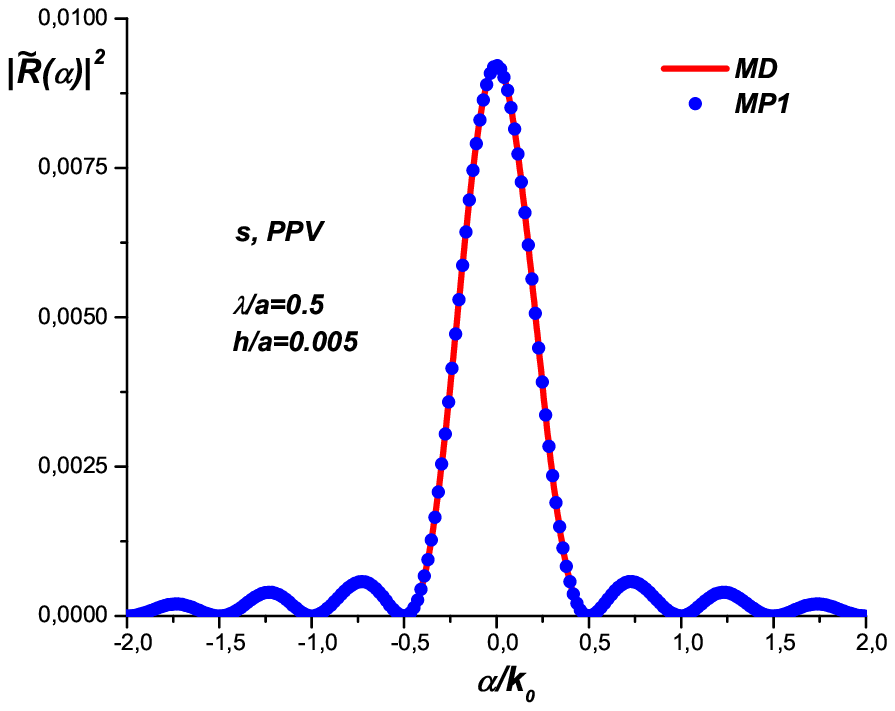}
\end{tabular}}
\caption{(Color online) Curves of $|\tilde R(\alpha)|^2$ vs $\alpha/k_0$ obtained with the direct method (continuous curve) and the first order perturbative method (circles) for a rectangular protuberance of width $a$ and height $h/a=0.005$ illuminated at normal incidence ($\theta_{0}=0^\circ$) in $s$ polarization. The wavelength is $\lambda/a=0.5$. The media are NPV (above) or PPV (below) with the same module of the refraction index. }
\label{figcompa1h1s}
\end{figure}

\begin{figure}
\resizebox{0.5\textwidth}{!}{
\begin{tabular}{c}
\includegraphics{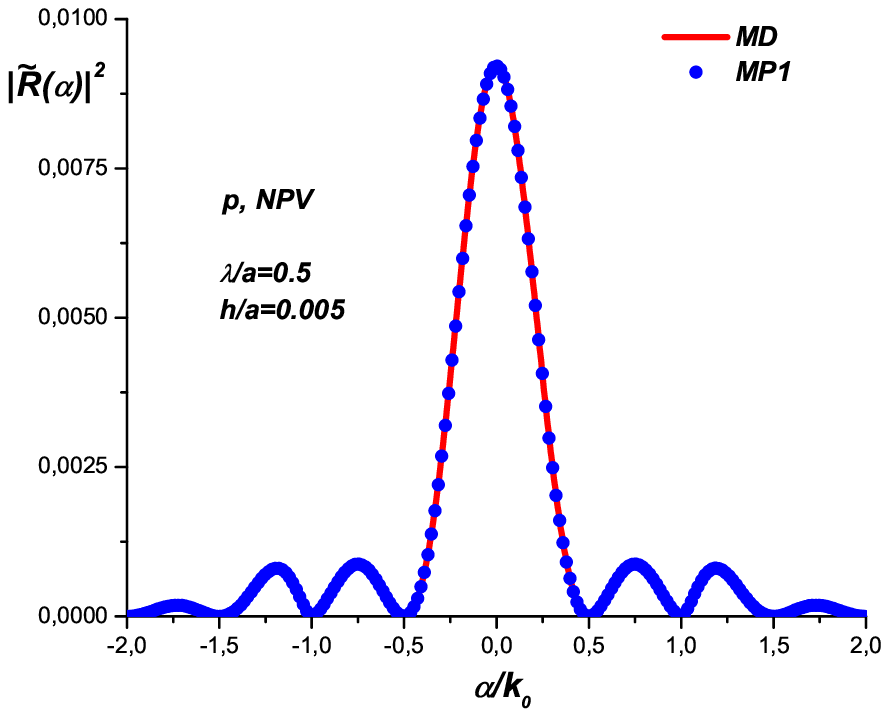}\\
\includegraphics{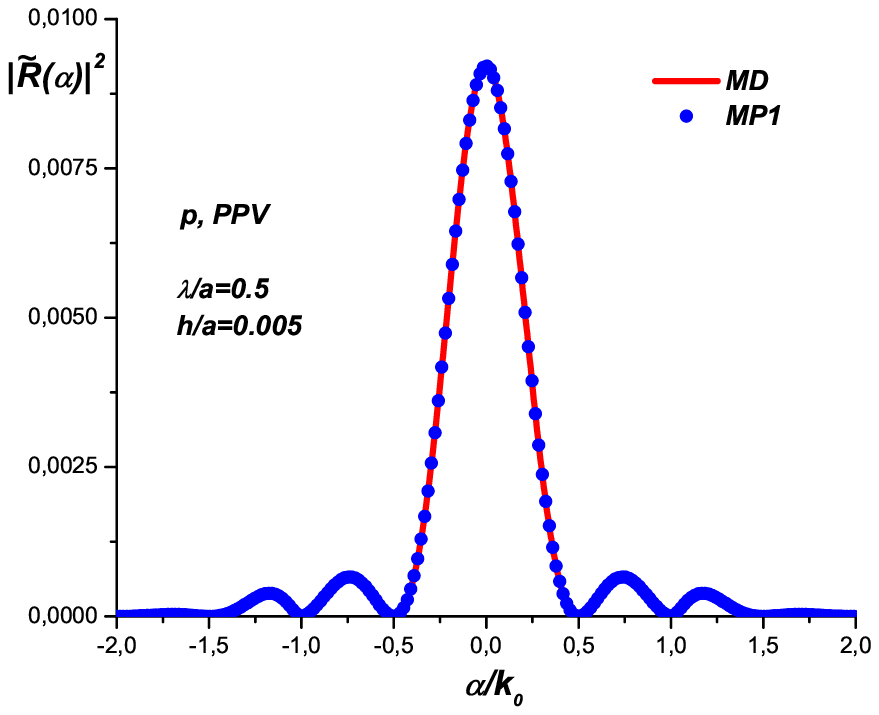}
\end{tabular}}
\caption{(Color online) Curves of $|\tilde R(\alpha)|^2$ vs $\alpha/k_0$ obtained with the direct method (continuous curve) and the first order perturbative method (circles) for a rectangular protuberance of width $a$ and height $h/a=0.005$ illuminated at normal incidence ($\theta_{0}=0^\circ$) in $p$ polarization. The wavelength is $\lambda/a=0.5$. The media are NPV (above) or PPV (below) with the same module of the refraction index. }
\label{figcompa1h1p}       
\end{figure}

To evaluate the reliability of the application of Ray\-leigh methods to non periodic rough surfaces with negative refractive index, we consider in this section the case of a rectangular protuberance (width $a$ and height $h$), illuminated at normal incidence ($\theta_{0}=0^\circ$). In this case, $g(x)=h\,\mbox{rec}(x/a)$, where $\mbox{rec}(u)$ is the rectangular function centered at the origin with unit width and height. 
In Figures \ref{figcompa1h1s} (for $s$ polarization) and \ref{figcompa1h1p} (for $p$ polarization) 
we compare the curves $|\tilde R(\alpha)|^2$ vs $\alpha/k_0$ obtained with the direct method (continuous curve) and the first-order perturbation method (circles) for the case $h/a=0.005$, $\lambda/a=0.5$ and for PPV and NPV media. 
We observe that both methods give an excellent agreement and that in all the cases they predict the presence of a principal maximum centered at the value of the spectral variable $\alpha=\alpha_{0}$, that corresponds to the specular reflection direction. This maximum, of width $2\lambda/a$, is surrounded by secondary maxima of width $\lambda/a$, in total coincidence with the results predicted by the scalar theory of diffraction for a slit $2\lambda$ wide. The power conservation criterion \ref{conservation1} is satisfied with an absolute error less than $0.001$. 
These results show that when the height of the protuberance is small, the perturbative method converges rapidly and coincides with the direct numerical method, as it has been already observed for periodical corrugations in NPV media \cite{rdep1}. We have verified that the same occurs for other values of geometric and incidence parameters and for other protuberance shapes. 

\begin{figure}
\resizebox{0.5\textwidth}{!}{
\begin{tabular}{c}
\includegraphics{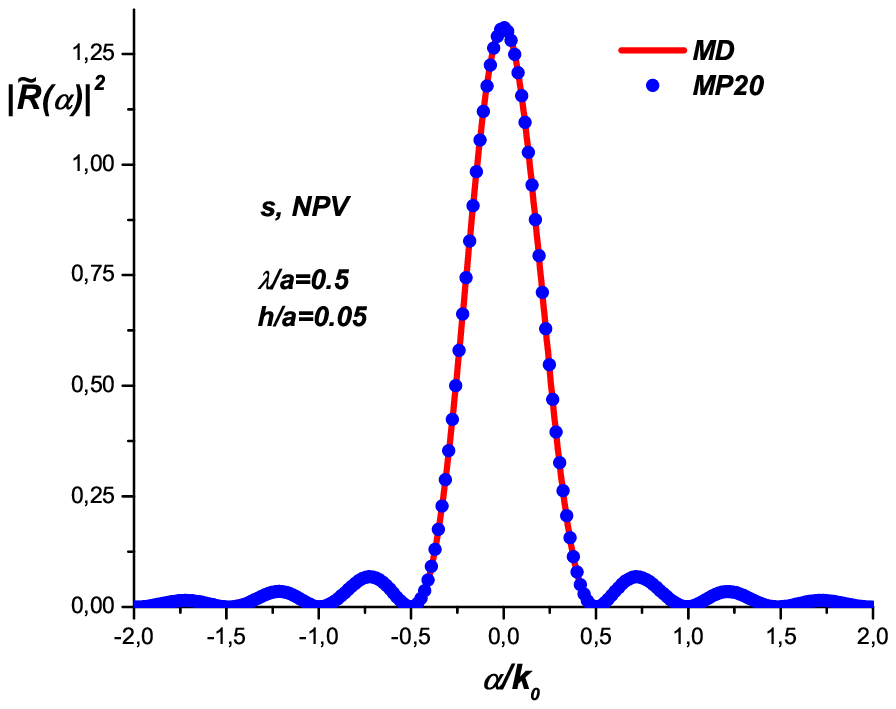}\\
\includegraphics{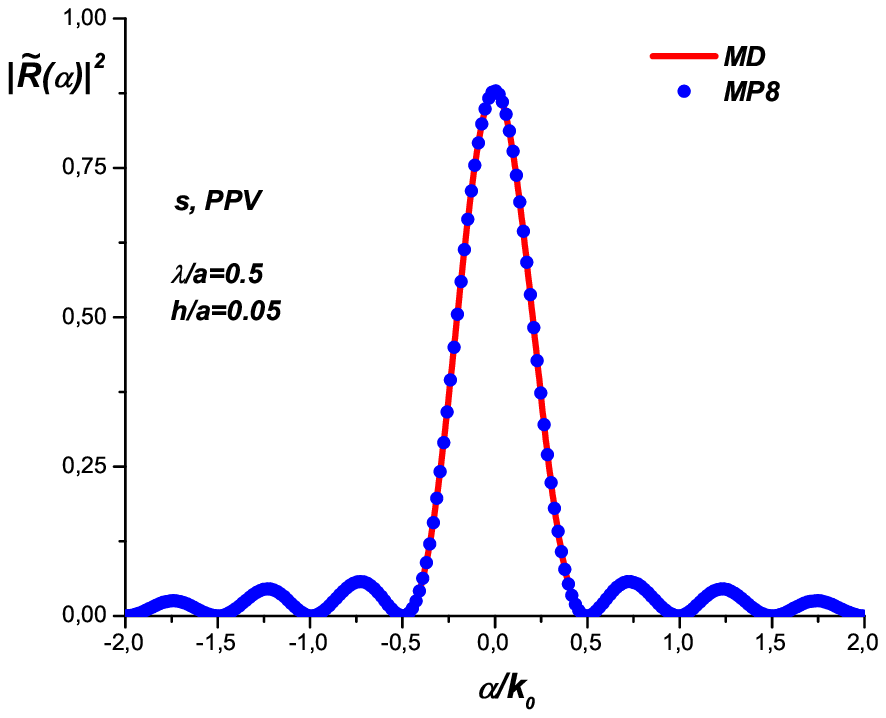}
\end{tabular}}
\caption{(Color online) Curves of $|\tilde R(\alpha)|^2$ vs $\alpha/k_0$ obtained with the direct method (continuous curve) and the first order perturbative method (circles) for a rectangular protuberance of width $a$ and height $h/a=0.05$ illuminated at  normal incidence ($\theta_{0}=0^\circ$) in $s$ polarization. The wavelength is $\lambda/a=0.5$. The media are NPV (above) or PPV (below) with the same module of the refraction index. }
\label{figcompa1h2s}       
\end{figure}

\begin{figure}
\resizebox{0.5\textwidth}{!}{
\begin{tabular}{c}
\includegraphics{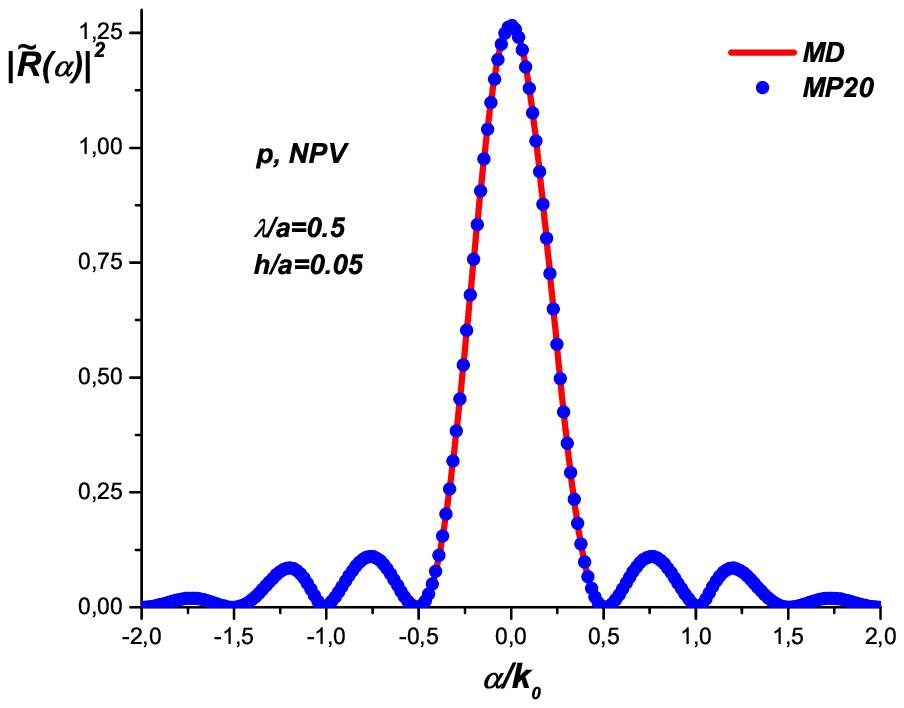}\\
\includegraphics{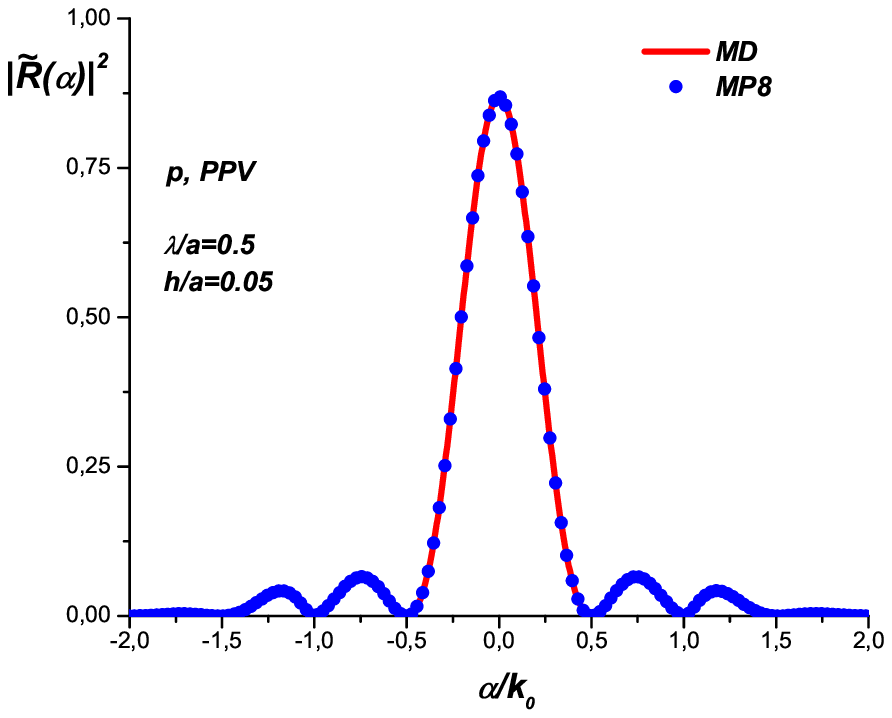}
\end{tabular}}
\caption{(Color online) Curves of $|\tilde R(\alpha)|^2$ vs $\alpha/k_0$ obtained width the direct method (continuous curve) and the perturbative method's first order (circles) for one rectangular protuberance of wide $a$ and height $h/a=0.05$ illuminated at normal incidence ($\theta_{0}=0^\circ$) in $p$ polarization. The wavelength is $\lambda/a=0.5$. The media considered are NPV (above) or PPV (below) with the same module of the refraction index. }
\label{figcompa1h2p}       
\end{figure}

In Figures \ref{figcompa1h2s} ($s$ polarization) and \ref{figcompa1h2p} ($p$ polarization) we repeat the comparison between the direct (continuous curve) and the perturbative (circles) methods, for the same situations considered 
in Figures \ref{figcompa1h1s} and \ref{figcompa1h1p}, except that now the height of the protuberance is 10 times higher ($h/a=0.05$). It is interesting to observe that in this case the perturbative method converges better in the PPV case, where the coincidence with the direct method is obtained in both polarizations in the $8^{th}$ perturbative order, while in the NPV case the coincidence is obtained in the $20^{th}$ perturbative order. 
For this protuberance height the power conservation criterion \ref{conservation1} is satisfied with an absolute error less than $0.07$ (PPV) or less than $0.13$ (NPV). 
We have found that for greater heights and for NPV media the perturbative method can fail to converge, while 
it still converges when the medium is PPV. For example, for a sinusoidal protuberance of the form $g(x)=\frac{h}{2} \lbrack 1+\cos(\frac{2\pi}{a}x)\rbrack$ rec$(x/a)$, $h/a=0.1$, $\lambda/a=0.5$, $\theta_{0}=0^\circ$ and PPV media, the perturbative results converge to those obtained with the direct numerical method in the $10^{th}$ order, although convergence is not obtained when the medium is NPV. 

\section{Results}
Having checked the reliability of the results obtained with the Rayleigh methods, we now turn our attention to the reflectivity of metamaterial protuberances whose indices of refraction have opposite signs. 
The normalized angular distribution of power scattered into medium 1 is given by the integrand in eq. (\ref{conservation2})
\begin{equation} 
\frac{dP^{(1)}}{d\alpha}=\frac{1}{2\pi a}\Re \, \frac{\beta_{\alpha}^{(1)}}{\beta_{0}^{(1)}}\, |\tilde{R}(\alpha)|^2 \,,
\label{potencia1}
\end{equation}
an expression which shows that only the values of $R(\alpha)$ in the radiative zone, $|\alpha/k_0| < \nu_{1}$, contribute to the scattered power. In this spectral zone the integrand in eq. (\ref{ecu3}) represents plane waves propagating away from the surface along a direction that forms a scattering angle $\theta_{s1}$, $\left(|\theta_{s1}|<\pi/2\right)$, with the $+y$ axis. 
\begin{figure}
\resizebox{0.5\textwidth}{!}{
\includegraphics{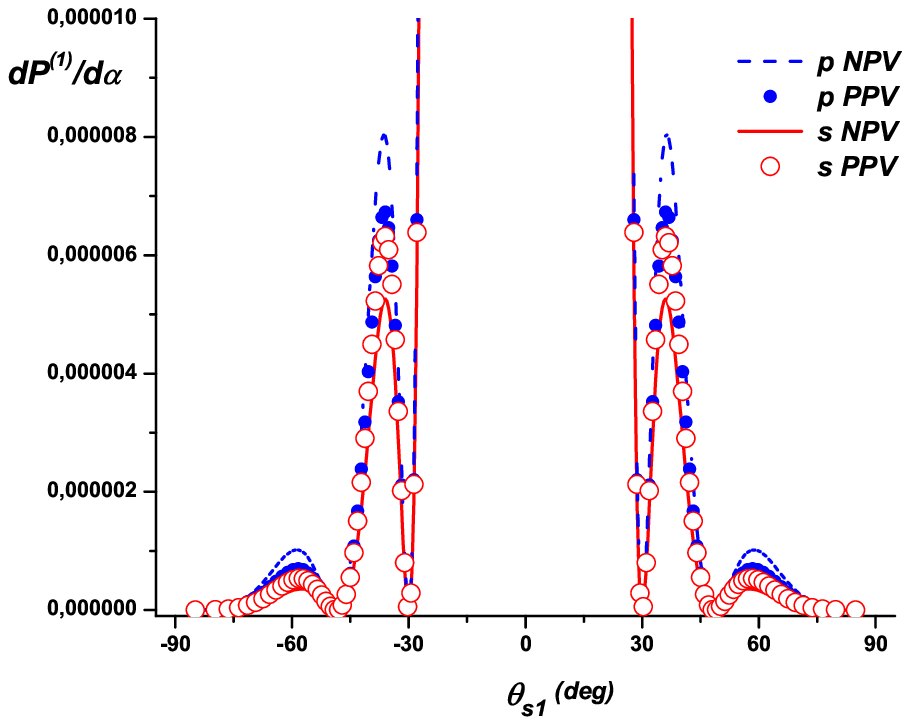}
}
\caption{(Color online) Angular distribution of power scattered into medium 1 for a sinusoidal protuberance. The incidence  parameters correspond to those of Figures \ref{figcompa1h1s} and \ref{figcompa1h1p} and the geometric parameters are $\lambda/a=0.25$, $h/a=0.0025$. 
}\label{potreflejada1}
\end{figure}

\begin{figure}
\resizebox{0.5\textwidth}{!}
{\includegraphics{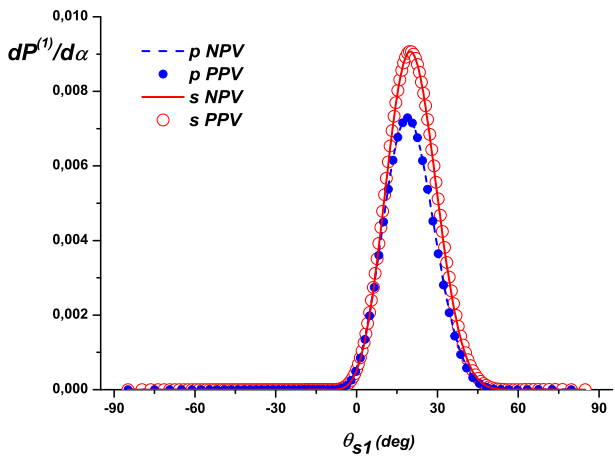}}
\caption{(Color online) Angular distribution of power scattered into medium 1 for a sinusoidal protuberance. The geometric and incident parameters correspond to those of Figure \ref{potreflejada1}, except that now $\theta_{0}=20^\circ$. 
}\label{potreflejada2a}
\end{figure}

\begin{figure}
\resizebox{0.5\textwidth}{!}
{\includegraphics{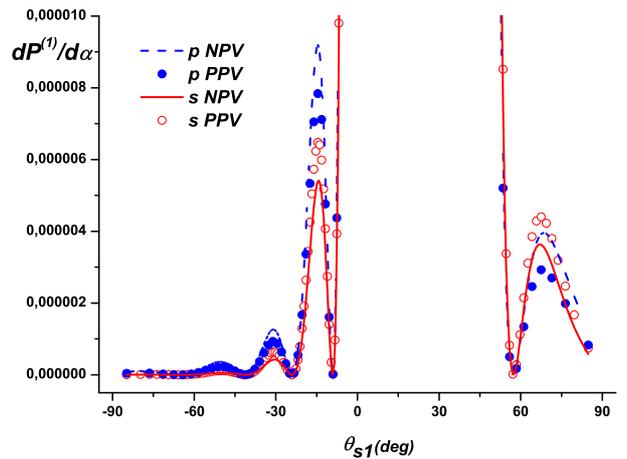}}
\caption{(Color online) Same as in Figure \ref{potreflejada2a}, but with a different scale to evidence the differences between curves in observation directions far away from the specular direction.}
\label{potreflejada2b}
\end{figure}

In Figure \ref{potreflejada1} we compare the angular distributions of reflected power that correspond to a single sinusoidal protuberance illuminated under normal incidence with 
$\lambda/a=0.25$ and $h/a=0.0025$. 
We have verified that the power conservation criterion is satisfied to an error less than $0.2\%$.
As in the examples considered in Figures \ref{figcompa1h1s} and \ref{figcompa1h1p}, 
close to the specular direction the four curves ($s$ PPV, $s$ NPV, $p$ PPV, $p$ NPV) show essentially the same response and 
so we have chosen an scale that amplifies the differences between the curves in observation directions $\theta_{s1} \neq \theta_{0}$. The first perturbative order exhibits analytically the coincidence of the responses in the specular direction, as can be seen from the expression (\ref{pertur}). From a physical point of view, such coincidence can be understood taking into account the low height considered in this example and the fact that for a flat surface: i) the $s$ and $p$ polarizations are indistinguishable in normal incidence and ii) the exchange between PPV and NPV refractive media changes the phase but not the absolute value of the Fresnel coefficients for the amplitudes of the reflected fields. 
However, it should be noted that despite the low height considered, for observation directions $\theta_{s1}$ away from $\theta_{0}$, the angular distribution of the reflected power corresponding to each incident polarization is sensitive to the sign of the refractive index of the metamaterial.
Similar features are observed for oblique incidences, as shown in Figures \ref{potreflejada2a} and \ref{potreflejada2b}, obtained for the same parameters used in Figure \ref{potreflejada1}, except that now $\theta_{0}=20^\circ$. 
%
\begin{figure}
\resizebox{0.5\textwidth}{!}
{\includegraphics{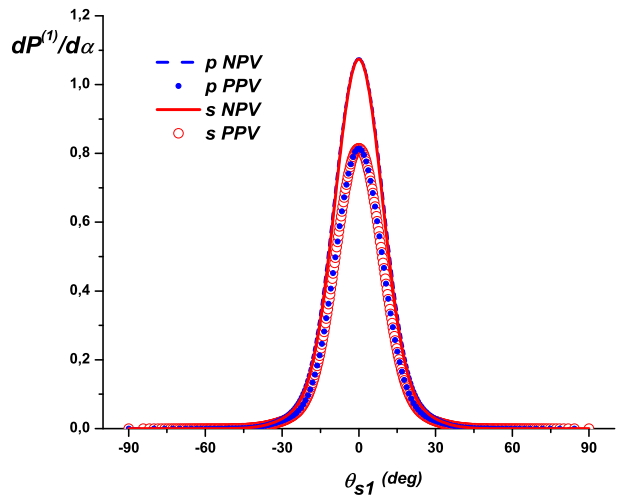}}
\caption{(Color online) Same as in Figure \ref{potreflejada1}, except that now $h/a=0.025$. 
}\label{potreflejada3}
\end{figure}
\begin{figure}
\resizebox{0.5\textwidth}{!}
{\includegraphics{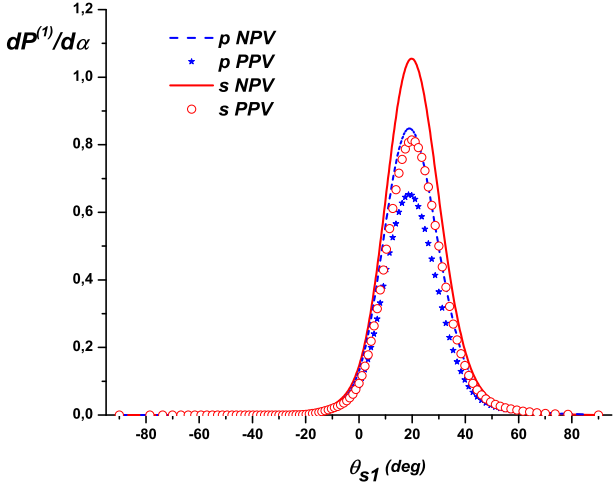}}
\caption{(Color online) Same as in Figure \ref{potreflejada3}, except that now $\theta_{0}=20^\circ$.  
}\label{potreflejada4}
\end{figure}
\begin{figure}
\resizebox{0.5\textwidth}{!}
{\includegraphics{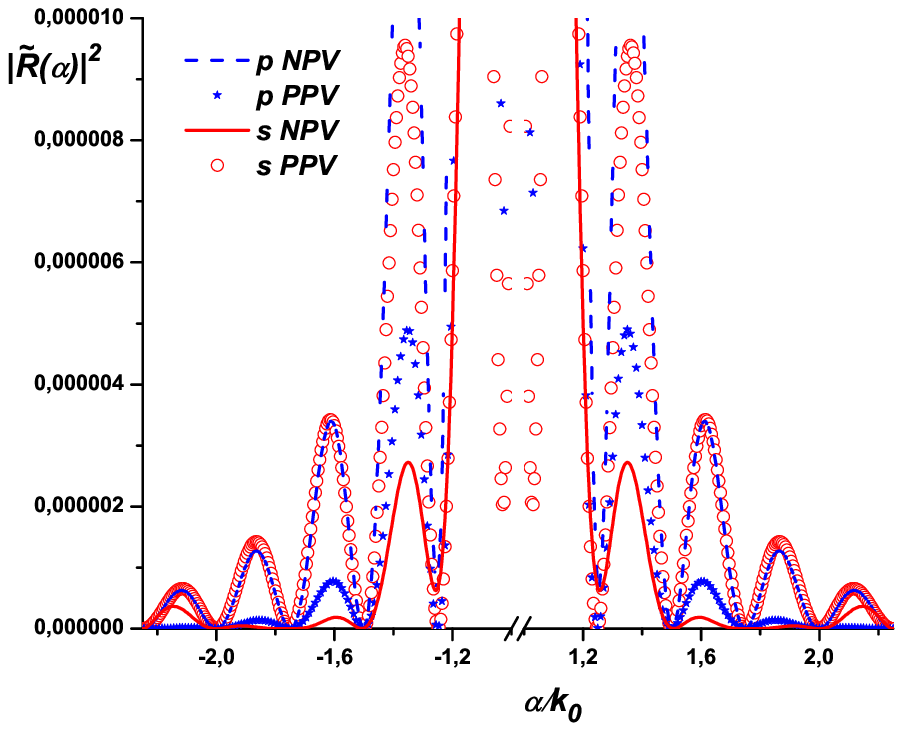}}
\caption{(Color online) Curves of $|\tilde R(\alpha)|^2$ vs $\alpha/k_0$ for a single sinusoidal protuberance. 
The geometric and incident parameters correspond to those of Figures \ref{potreflejada3}.}
\label{alfas}
\end{figure}

In Figures \ref{potreflejada3} (for normal incidence) and \ref{potreflejada4} (for $\theta_{0}=20^\circ$) we show the results corresponding to a single sinusoidal protuberance with a height value ten times greater than the value used in Figures \ref{potreflejada1}, \ref{potreflejada2a} and \ref{potreflejada2b}. We observe that when the protuberance height increases, the power scattered near the specular direction ($\theta_{s1} \approx \theta_{0}$) becomes more sensitive to the PPV-NPV interchange. 
%

The results presented in Figures \ref{potreflejada1}, \ref{potreflejada2a}, \ref{potreflejada2b}, \ref{potreflejada3} and \ref{potreflejada4} confirm the conjecture suggested by the conjugation symmetry, that is, the optical response along observation directions away from the specular direction could be used as a 
far-field indicator of the PPV/NPV character of shallowly corrugated metamaterial surfaces. Note that this indicator involves only the radiative range of the quantity 
$\tilde{R}(\alpha)$, i.e., the range in which $\tilde{R}(\alpha)$ represents the amplitude of propagating plane waves. However, in applications where near fields are involved, the non-radiative range of the quantity $\tilde{R}(\alpha)$ can also play an important role as a PPV/NPV indicator. This is due to the fact that in the non-radiative range the quantity $\tilde{R}(\alpha)$ represents evanescent waves that only affect the value of the fields near the surface and it is well known that the behavior of  evanescent waves changes dramatically depending on the sign of the refractive index of a medium. 
To explore this possibility, in Figure \ref{alfas} we compare the curves of $|\tilde{R}(\alpha)|^2$ vs $\alpha/k_{0}$ in the non-radiative zone for PPV and NPV corrugated metamaterials with the same geometric and incident parameters considered in Figures \ref{potreflejada3}.
It can be seen that, although the surface is optically almost flat ($h/\lambda=0.1$), the amplitudes of the evanescent fields generated during the scattering process are sensitive to the change of sign of the refractive index.



\section{Conclusion}

We have extended two electromagnetic scattering formalisms, originally developed for conventional (nonmagnetic) materials, to the case of corrugated materials with arbitrary (positive or negative) constitutive parameters. 
We are planning to use this extension to study the electromagnetic scattering from rough surfaces in applications motivated by the recent emergence of metamaterials with negative refractive index. We have presented examples showing that, 
even though they require different numerical treatments, both formalisms give coincident results for shallow corrugations (the known range of validity of Rayleigh methods) and that such results are in agreement with the predictions of physical optics. We have used these formalisms to illustrate the changes produced in the scattering properties of a surface with a protuberance when 
only the sign of the refractive index of the metamaterial below the surface is changed. To be realistic, we have considered lossy metamaterials and have centered our attention on the scattered fields reflected into the medium of incidence. In a next step, we are planning to consider the ideal case of lossless metamaterials in order to study the behavior of the scattered fields transmitted into the medium below the surface. 
Starting from a perfectly flat surface whose reflectivity is unaffected by the PPV/NPV transformation, 
we found that the angular distribution of scattered power corresponding to the same surface with a shallow added corrugation is more sensitive to the PPV/NPV interchange when the observation is not close to the specular direction and that the sensitivity  increases when the height of the corrugations increases. 
We have also shown that the near field can be highly affected by the PPV/NPV interchange, even for surfaces with very low height protuberances. \\

We acknowledge financial support from Consejo Nacional de Investigaciones Cient\'{\i}ficas 
y T\'ecnicas (CONICET), Agencia Nacional de Promoci\'on Cient\'{\i}fica y Tecnol\'ogica 
(BID 1728/OC--AR PICT-11--1785) and Universidad de Buenos Aires (UBA). 

%
%
%
%

\end{document}